# Combined Human, Antenna Orientation in Elevation Direction and Ground Effect on RSSI in Wireless Sensor Networks

Syed Hassan Ahmed , Safdar H. Bouk,  N. Javaid
Department of Electrical Engineering,
Comsats Institute of Information Technology,
Islamabad, Pakistan.
*Sani9585@gmail.com, {bouk,
nadeemjavaid}@comsats.edu.pk*

Iwao Sasase
Department of Information and Computer Science
Keio University,
Yokohama, Japan.
*sasase@ics.keio.ac.jp*

*Abstract*— In this paper, we experimentally investigate the combined effect of human, antenna orientation in elevation direction and the ground effect on the Received Signal Strength Indicator (RSSI) parameter in the Wireless Sensor Network (WSN). In experiment, we use MICAz motes and consider different scenarios where antenna of the transmitter node is tilted in elevation direction. The motes were placed on the ground to take into account the ground effect on the RSSI. The effect of one, two and four persons on the RSSI is recorded. For one and two persons, different walking paces e.g. slow, medium and fast pace, are analysed. However, in case of four persons, random movement is carried out between the pair of motes. The experimental results show that some antenna orientation angles have drastic effect on the RSSI, even without any human activity. The fluctuation count and range of RSSI in different scenarios with same walking pace are completely different. Therefore, an efficient human activity algorithm is need that effectively takes into count the antenna elevation and other parameters to accurately detect the human activity in the WSN deployment region.

## I. Introduction

Wireless Sensor Networks (WSN) has been investigated and researched extensively due to their large application domain. Recently, the parameter that has been considered in experiments of WSN deployment, localization, distance estimation etc is the variations in the received signal strength, called Received Signal Strength Indicator (RSSI). RSSI is an indication of power or power level present in the received signal strength at the receiving antenna. It is a very general feature or metric in most of the low-power radio enabled devices, i.e. Wireless Sensor Mote (WSM) [1-2].

In an ideal scenario, the power level of the received signal does not fluctuate. However, in practical scenario the RSSI is affected by different factors: e.g. physical distance, reflections of objects, environmental parameters, movement of objects or change in the environment, antenna position and polarization etc.

In this paper we experimentally analyze the combined effect of human activity, antenna orientation in elevation direction and ground effect on the RSSI of the MICAz WSM. The results show that even without human mobility, RSSI level is highly infected by the antenna orientation. The results also show that RSSI is immensely affected when all the above factors are collectively considered during the experiment. Remaining part of the paper is organized as follows: Previous work is discussed in Section-II. The experimental setup and results are explained in Section-III and IV, respectively. In last, Section-V concludes the paper.

## II. Previous Work

Several papers have been published in the past that investigate the effect of human or object movement, antenna orientation on RSSI or channel characteristics of WSM [3-9]. In [3], the channel propagation of MICAz mote with different antennas is analysed in the Anechoic Chamber and simulated as well. It has been observed that the radio pattern of the MICAz mote is not circular when a mote's antenna is tilted at orientations.

The authors in [4] have experimentally evaluated the impact of human movement on the RSSI in the indoor environment. The sensors were deployed inside the researchers' offices above the ground level and impact of human activity in investigated. It shows that there is significant effect of human activity on the RSSI of the mote.

In [5], the impact of antenna orientation on WSN performance has been experimentally investigated. It has been observed that when two motes' antennas are tilted in elevation direction in opposite direction has large effect on the RSSI compared to when two motes' antennas are parallel to each other. The signal propagation of T-mote Sky and MICAz mote is also investigated in [6].

The authors in [7], the sensor motes have been deployed on the ceiling where antenna is inverted and pointed to the floor and effect of human activity on the RSSI have measured.

The effect of human activity on RSSI of TelosB mote is experimentally investigated in [8]. The motes were placed at 1*m* to 3*m* high from the ground during the experiment. The human activity algorithm detects the presence of the human motion if the number of fluctuations in RSSI is less than the predefined threshold that is 60%. The RSSI fluctuations are also experimentally investigated in [9], where results of RSSI fluctuations of a moving node are compared with the accelerometer.

We note in the previous discussion that most of the previous work was focused on different factors that affect RSSI, i.e. antenna orientation and human movement. The collective influence of human activity, ground effect and antenna orientation on the received signal level are not investigated. In this paper, we propose an experimental approach to investigate the combined effect of human antenna orientation and ground effect on the RSSI of the MICAz motes.

## III. EXPERIMENTAL SETUP

The experimental setup to measure the combined effect of human activity, antenna orientation and ground effect is briefly discussed in this section. MICAz OEM Edition motes (MPR2600J) equipped with Atmega128L processor, CC2420 RF Chip and a half-wave external monopole antenna are used in this experiment [10]. The pair of MICAz motes is placed on the ground in a corridor of 5*m* wide at the distance of 3*m* apart. In the pair of motes, one mote works as a sender and other mote works as a receiver. The sender mote's antenna is tilted in elevation direction in reference to the receiver mote's antenna at varying angles, as shown in Figure 1.

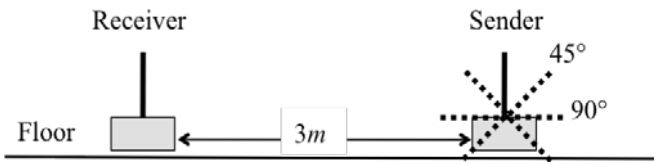

Figure 1. Experimental setup where a pair of MICAz motes is placed on the ground and antenna of *sender* mote is tilted in elevation direction with reference to the *receiver* mote.

RSSI readings of the successfully received packets are recoded for different scenarios, e.g. without human motion, with one and two persons moving between the nodes at slow, medium and fast pace and 4 persons moving at the medium pace.

## IV. EXPERIMENTAL RESULTS

The results of the experimental setup discussed in previous section are discussed in this section. Figure 2 shows the RSSI readings of the antenna orientation at different angles and without the human movement. It is observed in that illustration that signals strength (RSSI) is highly affected when the antenna of the sender node is tilted in elevation direction and nodes are placed on the ground. The percentage of constant RSSI reading (Frequency %) is very wide for some angles e.g. RSSI values for 90° elevation angle lie between -75dbm to -85dbm.

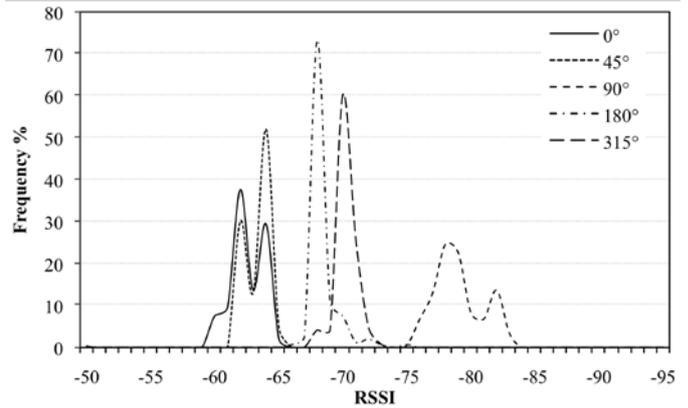

Figure 2. Frequency percentage vs RSSI at varying elevation angles with no movement

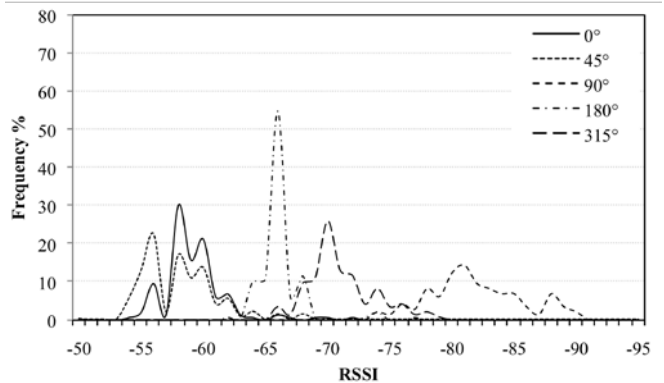

Figure 3. Frequency percentage vs RSSI at varying elevation angles with single person moving at slow pace

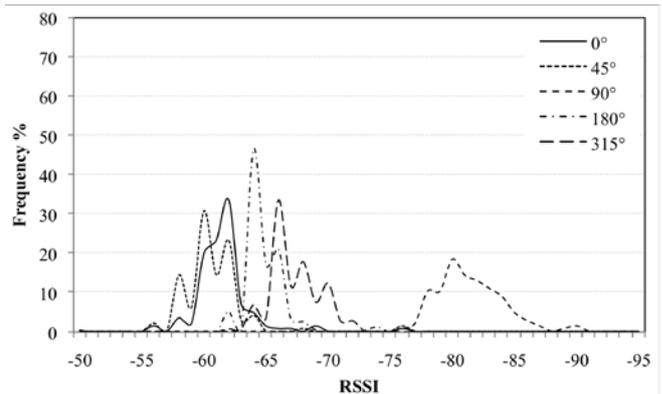

Figure 4. Frequency percentage vs RSSI at varying elevation angles with single person moving at medium pace

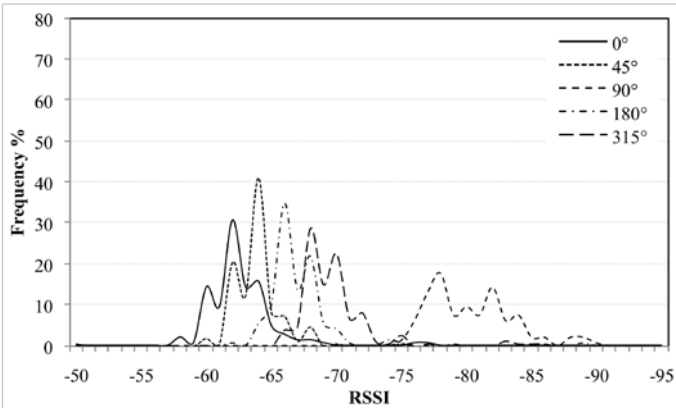

Figure 5. Frequency percentage vs RSSI at varying elevation angles with single person moving at fast pace

Figure 3, 4 and 5 show the effect of one person moving between nodes at slow, medium and fast pace, respectively. It is obvious from the figures that due to human mobility, frequency percentage of RSSI readings widens the range of RSSI values. The reason is that fluctuations in the RSSI values increase due to the human movement. If a person moves at slow, medium and fast pace, he/she obstructs the radio frequency signal for long, medium and short time respectively. This is evident from the figures that in case of slow pace movement, the RSSI value fluctuates between much wider range compared to the fast pace movement, refer Figure 3 and Figure 5.

The standard deviation of no movement, slow, medium and fast pace movement is shown in Figure 6. It shows that RSSI deviation is very less in case of no movement and deviation of slow movement is much higher for slower paces compared to faster paces at some angles where signal strength is stronger, i.e. 0°, 90° etc.

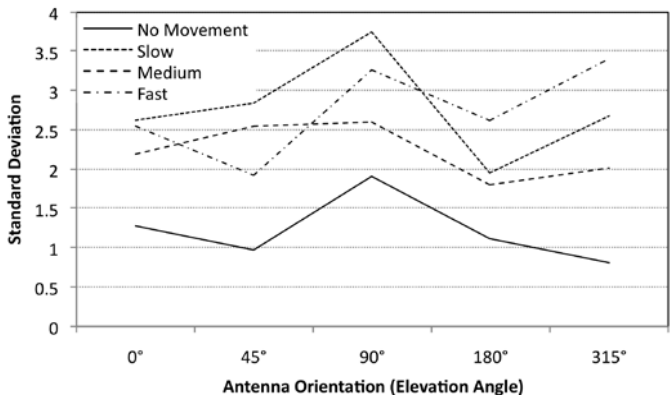

Figure 6. Standar deviation vs antenna orientation angles with single person moving at varying pace

The RSSI frequency ratios of a two persons slow and fast pace scenarios are shown in Figure 7 and Figure 8, respectively. It is obvious from those illustrations that the slower movement of two persons have relative high impact on the RSSI fluctuations compared to fast movement. The range of RSSI values is much wider in slow pace compared to the fast pace movement, refer Figure 7 and 8.

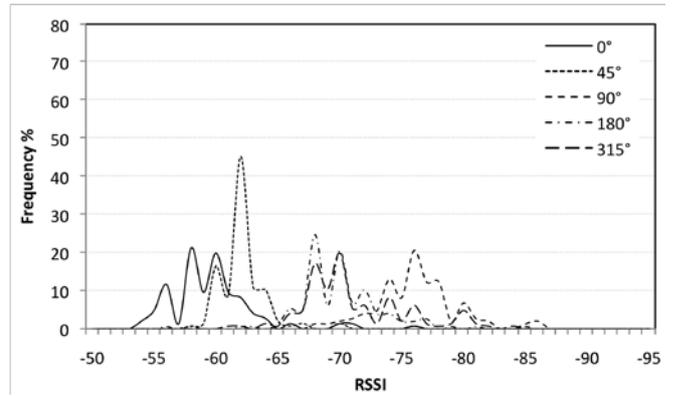

Figure 7. Frequency percentage vs RSSI at varying elevation angles with two persons moving at slow pace

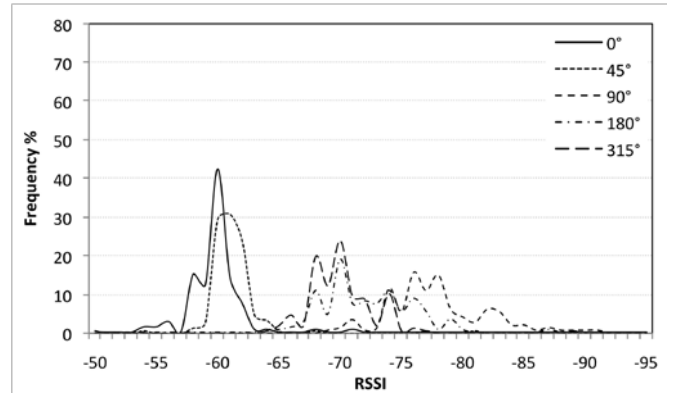

Figure 8. Frequency percentage vs RSSI at varying elevation angles with two persons moving at fast pace

The standard deviation of two person movement paces versus elevation angles is depicted in Figure 9. The RSSI deviation is with no movement is much smaller compared to the movement case. However, the interesting point in this figure and Figure 6 is that RSSI at 90° with no movement has larger deviation in no movement scenario compared to other angles.

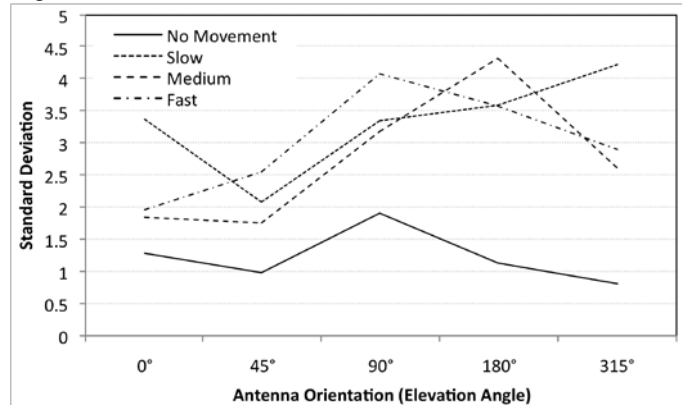

Figure 9. Standar deviation vs antenna orientation angles with two person moving at varying pace

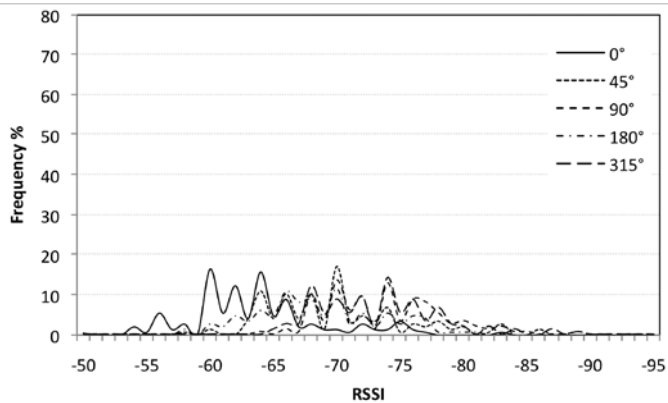

Figure 10. Frequency percentage vs RSSI at varying elevation angles with four persons moving at medium pace

Figure 10 shows RSSI frequency rate of the four persons medium pace movement case. It is observed that four persons movement between pair of nodes have really very high impact on the RSSI compared to the two and one-person slow pace movement scenarios, refer Figure 4 and 10.

It has been experimentally observed that antenna orientation, human movement with different movement paces and number of persons and ground effect have really high impact on the signal strength of the sensor mote radio frequency. In this case, it is very difficult to select a single RSSI fluctuation rate threshold point to detect the human activity in the WSN deployment area. Hence, the human activity detection algorithm based on the RSSI fluctuations must efficient and adaptive.

## V. CONCLUSIONS

In this paper, we experimentally examined the combined effect of human movement, antenna orientation in elevation direction and ground effect on the RSSI. We found that, even without the human movement, signal strength and its fluctuation are different at varying antenna angles. It is also observed that, human movement from slower to faster paces increases the fluctuations in the RSSI. Therefore, the human mobility detection in the WSN deployment area must be adaptive and consider all these factor to properly detect the human activity with minimum false counts.

## VI. ACKNOWLEDGMENT

This work was partly supported by Keio University 21st Century Centre of Excellence Program on "Optical and Electronic Device Technology for Access Network" and Fujitsu Laboratories.